\documentclass[aps,twocolumn,pra,amsmath,tightenlines]{revtex4-1}
\usepackage{epsfig,graphicx,times}
\usepackage{amstext}
\usepackage{amsmath}            %serve per le subequazioni
\usepackage{amssymb}            %serve per il simbolo "marchio registrato", \circledR
\usepackage{graphicx}           %serve per le figure eps, ps etcyy
\usepackage{mathrsfs}
\usepackage{bm}
\usepackage[tight]{subfigure}
\usepackage[colorlinks,linkcolor=blue,anchorcolor=blue,citecolor=blue,urlcolor=black]{hyperref} %urlcolor=black是链接的颜色，比如邮箱
\begin{document}

\title{Mechanical engineering of photon blockades in a cavity optomechanical system}

\author{Cuilu Zhai}
\affiliation{Key Laboratory of Low-Dimensional Quantum Structures and Quantum Control of Ministry of Education, Department of Physics and Synergetic Innovation Center for Quantum Effects and Applications, Hunan Normal University, Changsha 410081, China}

\author{Ran Huang}%
\affiliation{Key Laboratory of Low-Dimensional Quantum Structures and Quantum Control of Ministry of Education, Department of Physics and Synergetic Innovation Center for Quantum Effects and Applications, Hunan Normal University, Changsha 410081, China}%

\author{Baijun Li}%
\affiliation{Key Laboratory of Low-Dimensional Quantum Structures and Quantum Control of Ministry of Education, Department of Physics and Synergetic Innovation Center for Quantum Effects and Applications, Hunan Normal University, Changsha 410081, China}%

\author{Hui Jing}
\email{jinghui73@foxmail.com}
\affiliation{Key Laboratory of Low-Dimensional Quantum Structures and Quantum Control of Ministry of Education, Department of Physics and Synergetic Innovation Center for Quantum Effects and Applications, Hunan Normal University, Changsha 410081, China}%

\author{Le-Man Kuang}
\email{lmkuang@hunnu.edu.cn}
\affiliation{Key Laboratory of Low-Dimensional Quantum Structures and Quantum Control of Ministry of Education, Department of Physics and Synergetic Innovation Center for Quantum Effects and Applications, Hunan Normal University, Changsha 410081, China}%

\date{\today}

\begin{abstract}
We propose to mechanically control photon blockade (PB) in an optomechanical system with driving oscillators. We show that by tuning the mechanical driving parameters we achieve selective single-photon blockade (1PB) or two-photon blockade (2PB) as well as simultaneous 1PB and 2PB at the same frequency. This mechanical engineering of 1PB and 2PB can be understood from the anharmonic energy levels due to the modulation of the mechanical driving. In contrast to the optomechanical systems without any mechanical driving featuring PB only for specific optical detuning, our results can be useful for achieving novel photon sources with multi-frequency. Our work also opens up new route to mechanically engineer quantum states exhibiting highly nonclassical photon statistics.
%A mechanical pump in an optomechanical system (OMS) provides a way for controlling the optical response properties. However, it mainly focuses on the classical regime. Here, we present a method that engineers the purely quantum effects of single-photon blockade and two-photon blockade in an OMS with a mechanical pump. %We show the anharmonic energy-levels are further shifted in an optomechanical system with a mechanical pump.
%In addition; moreover
%We show that this mechanical engineering of photon blockades can be understood from the modulation of the energy levels by tuning the strength of the mechanical pump field. For a practical implementation, photon blockades are desirable to achieve novel photon sources. Different from the photon sources based on the nonlinearity-induced photon blockades, frequencies of the photons are limited by the fixed strength of the nonlinearity, our strategy can realize photon sources with optional frequencies. This suggests applications in quantum states engineering for the controllable generation of a train of single photons exhibiting highly nonclassical photon statistics.
%Two-photon and three-photon blockades in driven nonlinear systems
\end{abstract}

\maketitle

\section{\label{sec:level0}Introduction}
Achieving single-photon sources is highly desirable in modern quantum devices, including single-photon transistors~\cite{F.-Y. Hong2008}, quantum repeaters~\cite{Y. Han2010}, quantum-optical Josephson interferometer~\cite{D. Gerace2009}, as well as low-power sensors, qubit gates~\cite{H.-Z.Wu2010}, and non-classical light switches~\cite{K. Xia2018,S. Zhang2018,L. Tang2019}. Over the years, the studies and applications~\cite{A. Majumdar2013,D. E. Chang2007,G.W. Lin2015,X. Wang2016,X.-Y. Lü2015,X.-Y. Lü2013,I. Carusotto2009,M. J. Hartmann2010} of photon blockade (PB) open the possibility of realizing such goal originally proposed in a nonlinear cavity~\cite{Imamoglu1997}. We note that single-photon blockade (1PB)~\cite{L. Tian1992,Imamoglu1997}, the generation of a single photon in a nonlinear cavity can impede the probability of generating another photon in the cavity, has been experimentally demonstrated in different systems including cavity or circuit cavity quantum electrodynamics systems~\cite{K. M. Birnbaum2005,A. Reinhard2012,A. Faraon2008,C. Lang2011,A. J. Hoffman2011,K. Muller2015} and cavity-free devices~\cite{T. Peyronel2012}. In a recent experiment~\cite{C. Hamsen2017}, two-photon blockade (2PB)~\cite{S. S. Shamailov2010,C. Hamsen2017,A. Miranowicz2013,A. Miranowicz2014,C. J. Zhu2017,G. H. Hovsepyan2014,W.-W. Deng2015} has also been demonstrated, opening a route for creating two-photon logic gates. PB requires large nonlinearities which turns out to be highly challenging in practice. However, recently, unconventional PB, even with weak nonlinearities, based on the destructive quantum interferences between different dissipative pathways was theoretically proposed~\cite{LiewandSavona2010,ArkaMajumdar2012,W. Zhang2014,X. W. Xu2014,O. Kyriienko2014,Y. H. Zhou2016,S. Ferretti2013,H. J. Carmichael1985,MotoakiBamba2011,H. Flayac2017,B. Sarma2017} and then  experimentally demonstrated~\cite{H. J. Snijders2018,C. Vaneph2018}.

In theoretical studies, PB has also been studied in optical waveguides\cite{D. E. Chang2008}, coupled cavities~\cite{M. J. Hartmann2006,A. D. Greentree2006,D. G. Angelakis2007}, circuit-QED~\cite{Y. X. Liu2014}, gain cavity~\cite{Y. H. Zhou2018}, spinning resonator~\cite{RanHuang2018} and optomechanical system (OMS)~\cite{P. Rabl2011,A. Nunnenkamp2011}. We note that in the past decade, cavity optomechanics~\cite{W. P. Bowen2016,M. Aspelmeyer2014,T. J. Kippenberg2008,M. Metcalfe2014} has significantly extended fundamental studies and practical applications of coherent light-matter interactions, such as optomechanically induced transparency~\cite{G. S. Agarwal2010,S. Weis2010,A. H. Safavi-Naeini2011}, ultrasensitive sensing~\cite{E. Gavartin2012,A. G. Krause2012}, storage and transduction of light signals~\cite{V. Fiore2011}, and the investigation of nonlinear dynamics~\cite{F. Marquardt2006}. In addition, analogous to PB~\cite{liaoquad,H. Wang2015,X.W. Xu2013}, phonon blockade~\cite{H. Seok2017,N. Didier2011,H. Xie2017,H. Xie2018,H. Q. Shi2018,L.-L. Zheng2019} has also been studied in the OMS, offering a way to study the nonclassicality, entanglement, and dimensionality of the blockaded phonon states.

In this work, we study mechanical engineering of PB in the OMS with a driven oscillator~\cite{bowen2017,38bowen2016}. This coherent driving of mechanical oscillator has been experimentally realized in the OMS by using Josephson phase qubits~\cite{OConnell2010}, microwave electrical driven~\cite{J.Bochmann2013}, and other time-varying weak forces, which provide new tools to control optomechanical devices in applications from precision metrology~\cite{T.D.Stowe1997} to tunable photonics~\cite{M.L.Povinelli2005,M.Notomi2006}. For example, in recent experiments, mechanical pump was used to break time-reversal symmetry for light propagation~\cite{Breakingsymmetry2018nature}, to observe cascaded optical transparency~\cite{LFan2015}, and to control spin-phonon coupling~\cite{A. Barfuss2015,I. Yeo2014}. %Nonlinear dynamics of an optomechanical system with a coherent mechanical pump: Second-order sideband generation
However, previous studies on the role of mechanical pump in an OMS have mainly focused on the classical regimes, e.g., control of transmission rates instead of quantum noises. Here, we study mechanical engineering of PB, a purely quantum effect. We find that, by tuning the strength of the mechanical pump, the multi-frequency PB can be achieved in OMS, which is distinct from previous studies featuring PB only for specific optical detuning. Our results open a new route to study mechanical engineering of purely quantum optomechanical effect, such as mechanical squeezing~\cite{H. Tan2013,A. Kronwald2013}, photon-phonon entanglement~\cite{C. Joshi2012,J. Li2018}.

The remainder of this article is organized as follows. Section~\ref{model and solutions} introduces the physical model under our consideration. By theoretically treating the weak-driving term in Hamiltonian as a perturbation, we diagonalize the Hamiltonian and derive the anharmonic energy levels of the system. Then we analytically and numerically calculate the optical correlations of the system and show the mechanical engineering of PB. Finally, Sec.~\ref{OUTLOOKsec} is a summary and conclusion.
\section{\label{model and solutions}model and solutions}\label{theoreticalmodelsec}
\begin{figure}
  \centering
  \includegraphics[width=8.5cm]{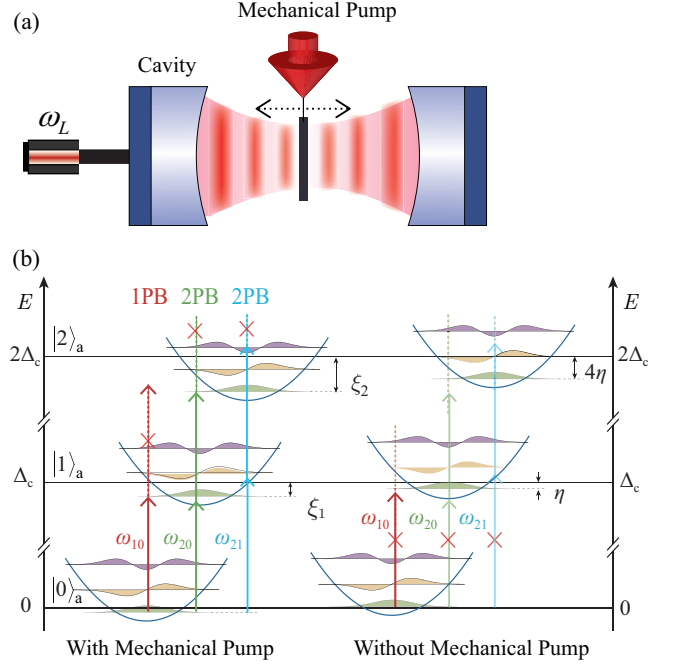}
  \caption{(a) Schematic of the OMS for a Fabry-Perot optical cavity with a moving mirror (mechanical resonator). A mechanical pump with strength $G$ is applied to the mechanical resonator. The mechanical pump was realized by the experimental setup of the cavity electro-optomechanical system consisting of a microtoroidal optomechanical oscillator with an integrated electrical interface that allows a radial force to be applied directly to the mechanical resonator as in Refs.~\cite{bowen2017,38bowen2016}. (b) Energy-level diagram of the OMS with (the left) and without mechanical pump (the right) for the relevant zero-photon state $\left|0\right\rangle_{a}$, one-photon state $\left|1\right\rangle_{a}$, and two-photon state $\left|2\right\rangle_{a}$.}\label{fp}
\end{figure}

We consider an OMS schematically illustrated in Fig.~\ref{fp}(a). The cavity is driven by a weak monochromatic laser field with frequency $\omega_{L}$. Meanwhile, a mechanical pump with strength $G$ is applied to excite the mechanical resonator. Moving to the rotating frame with respect to the driving laser field, the Hamiltonian of the system is of the form (hereafter $\hbar=1$) %Tunable multiphonon blockade in coupled nanomechanical resonators %Normal mode splitting and ground state cooling in a Fabry Perot optical cavity and transmission line resonator

\begin{eqnarray}
  H &=& H_{s}+H_{d}+H_{p},\label{Hamilton} \nonumber \\
  H_{s} &=& \Delta_{c}a^{\dagger}a+\omega_{m}b^{\dagger}b+g_{0}a^{\dagger}a(b^{\dagger}+b),\label{Hamitonianwithoutdriving} \nonumber \\
  H_{p} &=& G(\hat{b}^{\dagger}+\hat{b}),\nonumber \label{Hamiltonpump} \\
  H_{d} &=& \Omega(\hat{a}^{\dagger}+\hat{a}). \label{Hamiltondr}
\end{eqnarray}
Here, $a$ ($a^{\dagger}$) and $b$ ($b^{\dagger}$) are, respectively, the annihilation (creation) operators of the optical cavity field and the mechanical mode, with respective resonant frequencies $\omega_{c}$ and $\omega_{m}$. $g_{0}$ represents the single-photon coupling strength between the cavity field and the mechanical resonator. $\Delta_{c}=\omega_{c}-\omega_{L}$ is the detuning between the cavity mode and the driving field.
%(with frequency $\omega_{L}$).
Here, the strong-coupling regime, where the coupling rate $g_{0}$ exceeds the cavity amplitude decay rate $\gamma_c$ is required. $H_{s}$ is the Hamiltonian of the OMS without driving term and pumping term. The interaction between the mechanical mode and the pumping field is described as $ H_{p}$. The mechanical pump is used to excite phonons in the mechanical mode. The Hamiltonian of the mechanical pump was realized by the cavity electro-optomechanical system consisting of a microtoroidal optomechanical oscillator with an integrated electrical interface that allows a radial force to be applied directly to the mechanical resonator as in Refs.~\cite{bowen2017,38bowen2016}. $H_{d}$ describes the coupling between the cavity and the weak optical driving laser. The amplitude of the driving field $\Omega$ is related to the input laser power $P_{\text{in}}$ and cavity decay rate $\gamma_{c}$ by $\left|\Omega\right|=\sqrt{P_{\text{in}}\gamma_{c}/\omega_{L}}$.

Here, we show that the Hamiltonian exhibits an anharmonic energy-level configuration, which is crucial to realize 1PB and 2PB. In order to study the eigenenergies of the system,
we consider $\left|n\right\rangle _{a}$ and $\left|m\right\rangle _{b}$ ($n,m=0, 1, 2...$) as the
harmonic-osillator number states of the cavity field and
the mechanical mode, respectively. We consider a unitary transformation
\begin{equation}\label{D}
  D(n)=\text{exp}[(g_{0}n+G)/\omega_{m}(b-b^\dagger)],
\end{equation}
applied to $H_{s}+H_{p}$, where $n=a^{\dagger}a$. The Hamiltonian $\tilde{H}=D^{\dagger}(H_{s}+H_{p})D$ is generalized with the form
\begin{equation}\label{tildeHamiltonian}
\tilde{H}=\Delta_{c}a^{\dagger}a+\omega_{m}b^{\dagger}b-\eta (a^{\dagger}a)^{2}-\delta a^{\dagger}a-\frac{G^{2}}{\omega_{m}}.
\end{equation}
Clearly, the Hamiltonian $\tilde{H}$ satisfies
\begin{equation}\label{tildeHeigenstates}
\tilde{H}\vert n\rangle _{a}\vert m\rangle _{b}=E_{nm}\vert n\rangle _{a}\vert m\rangle _{b},
\end{equation}
where the eigenvalues are
\begin{equation}\label{E(mn)}
E_{nm}=n\Delta_{c}+m\omega_{m}-n^{2}\eta-n \delta-\frac{G^{2}}{\omega_{m}},
\end{equation}
where, $\eta=g^{2}_{0}/\omega_{m}$ and $\delta=2g_{0}G/\omega_{m}$. We multiply the operator $D(n)$ from the two sides of Eq.~(\ref{tildeHeigenstates}), one can obtain
\begin{equation}\label{Hs+Hpeigenstates}
(H_{s}+H_{p})\vert n\rangle _{a}\vert\tilde{m}(n)\rangle_{b}=E_{nm}\vert n\rangle _{a}\vert \tilde{m}(n)\rangle_{b}.
\end{equation}
The $n$-photon displaced number states in Eq.~(\ref{Hs+Hpeigenstates}) are
defined by
\begin{equation}\label{Hmi}
\vert \tilde{m}(n)\rangle_{b}
=D(n)\vert m\rangle_{b}.
\end{equation}
Especially, $\left|\tilde{m}(0)\right\rangle_{b}=\text{exp}[(G/\omega_{m})(b-b^{\dagger})]\left|m\right\rangle_{b}$. From the Eq.~(\ref{E(mn)}), we can know that anharmonic energy levels of the system are obtained based on the nonlinear coupling and the mechanical pump. We note that the energy frequency shift with $n^{2}\eta$ in Eq.~(\ref{E(mn)}) is caused by the nonlinear optomechanical interaction,
which has been studied in previous literature~\cite{P. Rabl2011}. Due to the mechanical pump, the energies can be modulated by the terms of $n \delta$ and $G^{2}/\omega_{m}$.

Since the optical driving strength is much smaller than the cavity decay
rate, $\Omega\ll\gamma_{c}$, only the lower energy states $\vert0\rangle _{a}$,
$\vert1\rangle _{a}$, and $\vert2\rangle_{a}$ of the
cavity field are occupied. For convenience, the eigen spectrum of the Hamiltonian $H_{s}+H_{p}$ limited in the zero-, one-, and two-photon cases is shown in Fig.~\ref{fp}(b).
%Thus the anharmonic energy-level structure of the system is produced.
The nonlinear resonator exhibits the energy shifts%Single-photon nonlinearities in a strongly driven optomechanical system with quadratic coupling
\begin{eqnarray}
\xi_{n}&=&\frac{(n^{2}g^{2}_{0}+2ng_{0}G+G^{2})}{\omega_{m}},\label{xi2}
\end{eqnarray}
in $n$-photon states without phonon sidebands respectively. Without the mechanical pump, i.e., $G=0$, the anharmonicity reduces to $\xi_{n}=n^{2}g^{2}_{0}/\omega_{m}$. When $G\not=0$, this energy shift can be modulated by tuning the strength of the mechanical pump filed, which can be used to realize photon sources with optional frequencies.

In Fig.~\ref{fp}(b), for the input laser frequency $\omega_{L}=\omega_{10}=\omega_{c}-2\eta$, no
PB can emerge without mechanical pump. However, for the same driving laser, 1PB can be realized with the mechanical pumping strength $G =g_{0}/2$. Similarly, with the mechanical pump, 2PB corresponding to the transitions $\vert 0\rangle_{a}\vert \tilde{0}(0)\rangle_{b} \rightarrow\vert 2\rangle_{a}\vert \tilde{0}(0)\rangle_{b}$ and $\vert 0\rangle_{a}\vert \tilde{0}(0)\rangle_{b} \rightarrow\vert2\rangle_{a}\vert \tilde{2}(2)\rangle_{b}$  can occur for the driving frequency $\omega_{20}$ and $\omega_{21}$, respectively, but can not emerge in the system without the mechanical pump for the same driving frequency. In the OMS without the mechanical pump, 1PB or 2PB occurs at particular optical driving frequency, which fulfills the single-photon or two-photon resonance transition condition. However, by tuning the strength of the mechanical pump, PBs can be realized with optional frequencies. This is a clear signature of mechanical engineering of PBs, which opens up a new route to achieve single-photon or few-photon sources with multi-frequency.

%\subsection{\label{sec:level2} Mechanical engineering of PB}\label{Photonblockadesec}
Next, we analytically calculate the second-order and the third-order correlation functions of cavity photons by treating the weak-driving term for Hamiltonian (\ref{Hamiltondr}) as a perturbation.
For the sufficient small $\Omega$, only the lower energy levels of the system are excited. Then the general state of the system in the few-photon subspace can be written as
\begin{eqnarray}\label{eq:psi}
\vert \psi(t)\rangle&=&\sum_{n=0}^{n=3}\sum_{m=0}^{\infty}C_{n,m}(t)\vert n\rangle_{a}\vert \tilde{m}(n)\rangle_{b},
\end{eqnarray}
where coefficients $C_{n,m}$ describe the probability amplitudes of the corresponding states respectively. The single-photon, two-photon, and three-photon displaced number states for the mechanical modes can be obtained from Eq.~(\ref{Hmi}) and read
\begin{eqnarray}\label{photon displaced number states}
  \vert \tilde{m}(1)\rangle_{b} &=& \text{exp}[\frac{g_{0}}{\omega_{m}}(b-b^{\dagger})+ \frac{G}{\omega_{m}}(b-b^{\dagger})]\vert m\rangle_{b},\nonumber \\
  \vert \tilde{m}(2)\rangle_{b} &=& \text{exp}[\frac{2g_{0}}{\omega_{m}}(b-b^{\dagger})+ \frac{G}{\omega_{m}}(b-b^{\dagger})]\vert m\rangle_{b},\nonumber \\
  \vert \tilde{m}(3)\rangle_{b} &=& \text{exp}[\frac{3g_{0}}{\omega_{m}}(b-b^{\dagger})+ \frac{G}{\omega_{m}}(b-b^{\dagger})]\vert m\rangle_{b}.
\end{eqnarray}

Considering the dissipation of the cavity mode (the time in the case of $1/\gamma_{c}\ll t \ll 1/\gamma_{m}$, $\gamma_{m}$ represents the mechanical decay), we phenomenologically add an anti-Hermitian term to Hamiltonian (\ref{Hamilton}) \cite{liaoquad}. The effective non-Hermitian Hamiltonian takes the form
\begin{equation}\label{eq:Hamiltonian}
H_{\mathrm{eff}}=H-i\frac{\gamma_{c}}{2}a^{\dagger}a.
\end{equation}
In terms of Eqs.~(\ref{eq:psi}) and (\ref{eq:Hamiltonian}), and the Schr\"{o}dinger equation $i d\psi(t)/dt=H_{\mathrm{eff}}\psi(t)$, we obtain the equations of motion for the probability amplitudes
\begin{eqnarray}\label{probability amplitudes motion}
  \dot{C}_{0,m} &=& -iE_{0,m}C_{0,m}-i\Omega\sum_{m^{\prime}=0}^{\infty}\, _{b}\!\langle \tilde{m}(0)\vert \tilde{m^{\prime}}(1)\rangle _{b}C_{1,m^{\prime}},\nonumber \\
  \dot{C}_{1,m} &=& -\Gamma_{1}C_{1,m}-i\Omega\sum_{m^{\prime}=0}^{\infty}\,_{b}\!\langle \tilde{m}(1)\vert\tilde{m^{\prime}}(0)\rangle _{b}C_{0,m^{\prime}}\notag \\
  &&-i\sqrt{2}\Omega\sum_{m^{\prime}=0}^{\infty}\,_{b}\!\langle \tilde{m}(1)\vert\tilde{m^{\prime}}(2)\rangle _{b}C_{2,m^{\prime}}, \nonumber \\
  \dot{C}_{2,m} &=& -\Gamma_{2}C_{2,m} -i\sqrt{2}\Omega\sum_{m^{\prime}=0}^{\infty}\,_{b}\!\langle \tilde{m}(2)\vert\tilde{m^\prime}(1)\rangle _{b}C_{1,m^{\prime}} \notag \\
  &&-i\sqrt{3}\Omega\sum_{m^{\prime}=0}^{\infty}\,_{b}\!\langle \tilde{m}(2)\vert\tilde{m^{\prime}}(3)\rangle _{b}C_{3,m^{\prime}},  \nonumber \\
 \dot{C}_{3,m}&=&-\Gamma_{3}C_{3,m}-i\sqrt{3}\Omega\sum_{m^{\prime}=0}^{\infty}\,_{b}\!\langle \tilde{m}(3)\vert\tilde{m^{\prime}}(2)\rangle _{b}C_{2,m^{\prime}}, \nonumber \\
\end{eqnarray}
where $ \Gamma_{n}=n\gamma_{c}/2+i E_{n,m}$. These transiton rates can be calculated using the relations $_{b}\!\langle \tilde{l}(n^{\prime})\vert \tilde{k}(n)\rangle _{b}= _{b}\!\langle l\vert D(n-n^{\prime})\vert k\rangle _{b}$ and
\begin{eqnarray}
_{b}\!\langle l\vert e^{\alpha(b^{\dagger}-b)}\vert k\rangle _{b} =\begin{cases}
\sqrt{\frac{l!}{k!}}e^{-\frac{\alpha^{2}}{2}}(-\alpha)^{k-l}L_{l}^{k-l}(\alpha^{2}), & k\geq l \nonumber \\
\sqrt{\frac{k!}{l!}}e^{-\frac{\alpha^{2}}{2}}(-\alpha)^{l-k}L_{k}^{l-k}(\alpha^{2}), & l>k
\end{cases}  \\
&
\end{eqnarray}
where $L_{r}^{s}(x)$ is generalized Laguerre polynomial.%Photon blockade in quadratically coupled optomechanical systems

In the weak-driving case, we have the following approximate formulas: $C_{0,m}\sim1$,
$C_{1,m}\sim\Omega/\gamma_{c}$, $C_{2,m}\sim\Omega^{2}/\gamma_{c}^{2}$, $C_{3,m}\sim\Omega^{3}/\gamma_{c}^{3}$. %When a weak driving field is applied to the cavity, it may excite a single photon or two photons into the cavity and the probability  of zero photon is probably about $1$, the probability of single photon is on the scale of $\Omega$ and $C_{2,m}$ is on the scale of $\Omega^{2}$. %%Photon blockade in quadratically coupled optomechanical systems
To approximately solve Eq.~(\ref{probability amplitudes motion}), we neglect the higher-order terms of $\Omega$ in the weak driving regime. Note this approximation has been widely utilized in cavity QED~\cite{Rebi2002,Leach2004} and OMS~\cite{liaoquad,Komar2013} for studying the photon statistics. %Phonon blockade in a quadratically coupled optomechanical system
For an initial empty cavity, we have $C_{1,m}(0)=0$, $C_{2,m}(0)=0$, $C_{3,m}(0)=0$; then the long-time solution of Eq.~(\ref{probability amplitudes motion}) can be approximately obtained as
\begin{eqnarray}\label{longterm solution}
  C_{0,m} &=& C_{0,m}(0)e^{-iE_{0,m}t}, \nonumber \\
  C_{1,m} &=& -\Omega\sum_{l=0}^{\infty}\frac{_{b}\!\langle \tilde{m}(1)\rangle\vert \tilde{l}(0)\rangle _{b}C_{0,l}(0)e^{-iE_{0,l}t}}{(E_{1,m}-E_{0,l}-i\frac{\gamma_{c}}{2})}, \nonumber \\
  C_{2,m} &=& \sqrt{2}\Omega^{2}\sum_{n,l=0}^{\infty}\frac{_{b}\!\langle \tilde{m}(2)\vert\tilde{n}(1)\rangle _{b}}{(E_{1,n}-E_{0,0}-i\frac{\gamma_{c}}{2})} \nonumber \\
  &&\times\frac{_{b}\!\langle \tilde{n}(1)\vert \tilde{l}(0)\rangle _{b}C_{0,l}(0)e^{-iE_{0,l}t}}{(E_{2,m}-E_{0,l}-i\gamma_{c})}, \nonumber \\
  C_{3,m} &=& -\sqrt{6}\Omega^{3}\sum_{m^{\prime},q,l=0}^{\infty}\frac{_{b}\!\langle \tilde{m}(3)\vert \tilde{m^{\prime}}(2)\rangle_{b}\,_{b} \!\langle\tilde{m^{\prime}}(2)\vert\tilde{q}(1)\rangle _{b}}{(E_{1,q}-E_{0,l}-i\frac{\gamma_{c}}{2})}\nonumber \\
  &&\times\frac{_{b}\!\langle \tilde{q}(1)\vert \tilde{l}(0)\rangle_{b} C_{0,l}(0)e^{-iE_{0,l}t}}{(E_{2,m^{\prime}}-E_{0,l}-i\gamma_{c})(E_{3,m}-E_{0,l}-i\frac{3\gamma_{c}}{2})},
  \nonumber \\
\end{eqnarray}
where $C_{0,m}(0)$ and $C_{0,l}(0)$ are determined by the initial state of the mechanical modes. %Photon blockade in quadratically coupled optomechanical systems
%\begin{eqnarray}
%\left.\left\langle \tilde{m}(1)\right|\tilde{m}^{\prime}\right\rangle _{b}&=&e^{-\alpha (\hat{b}^{\dagger}-\hat{b})}, \notag \\
%\left.\left\langle \tilde{m}(2)\right|\tilde{m}^{\prime}(1)\right\rangle _{b}&=&e^{-\alpha (\hat{b}^{\dagger}-\hat{b})}.
%\end{eqnarray}
%It's clear that $\left.\left\langle \tilde{m}(1)\right|\tilde{m}^{\prime}\right\rangle _{b}$ is equal to $\left.\left\langle \tilde{m}(2)\right|\tilde{m}^{\prime}(1)\right\rangle _{b}$, that make the analytical easier.
We assume that the membrane is initially in its ground state $\vert0\rangle _{b}$, i.e., $C_{0,m}(0)=\delta_{m,0}$. For simplicity, we consider the Taylor expansion of the unitary operators, then the long-time solutions of the system can be obtained from Eq.~(\ref{longterm solution}). % Photon blockade in a quadratically coupled optomechanical system
Accordingly, the probability of zero-photon, one-photon, two-photon and three-photon of the system can be obtained.

The equal-time second-order correlation and the equal-time third-order correlation can be written as $g^{(2)}(0)=2P_{2}/(P_{1}+2P_{2})^{2}$ and $g^{(3)}(0)=6P_{3}/(P_{1}+2P_{2}+3P_{3})^{3}$, i.e.,
\begin{eqnarray}
g^{(2)}(0)&=&\sum_{m=0}^{\infty}\frac{2\vert C_{2,m}\vert ^{2}}{(\vert C_{1,m} \vert ^{2}+2\vert C_{2,m}\vert ^{2})^{2}}, \label{g2_3}\\
g^{(3)}(0)&=&\sum_{m=0}^{\infty}\frac{6\vert C_{3,m}\vert ^{2}}{(\vert C_{1,m}\vert ^{2}+2\vert C_{2,m}\vert ^{2}+3\vert C_{3,m}\vert ^{2})^{3}}\label{g3_3},
\end{eqnarray}
where $P_{1}=\sum_{m=0}^{\infty}\left|C_{1,m}(t)\right|^{2}$, $P_{2}=\sum_{m=0}^{\infty}\left|C_{2,m}(t)\right|^{2}$ and $P_{3}=\sum_{m=0}^{\infty}\left|C_{3,m}(t)\right|^{2}$ are the probabilities for finding a single photon, two photons and three photons in the cavity, respectively. %Photon blockade in quadratically coupled optomechanical systems

In the weak-driving case, $P_{1}^{2}\gg P_{2}^{2}$ and $P_{2}^{2}\gg P_{3}^{2}$. Thus, correlation functions are reduced to
\begin{eqnarray}
g^{(2)}(0)&\approx&\frac{2P_{2}}{P_{1}^{2}}, \label{g2_2}\\
g^{(3)}(0)&\approx&\frac{6P_{3}}{P_{1}^{3}}\label{g3_2}.
\end{eqnarray}

We now turn to the numerical solution case. In fact, $g^{(2)}(0)=\langle a^{\dagger}a^{\dagger} aa \rangle /\langle a^{\dagger}a\rangle ^{2}$ and $g^{(3)}(0)=\langle a^{\dagger}a^{\dagger}a^{\dagger}aaa\rangle /\langle a^{\dagger}a\rangle ^{3}$~\cite{C. Hamsen2017}. The classical and quantum fluctuations of the environmental degrees of freedom will introduce damping to the cavity field and mechanical oscillator~\cite{Gardiner2004,Walls2008}, as required by the fluctuation-dissipation theorem~\cite{Kubo1966}. After taking into account both optical and mechanical dissipations, the dynamical evolutoin of the system is described by the master equation
\begin{eqnarray}\label{master equation}
  \dot{\rho}&=&i[\rho,H]+\frac{\gamma_{c}}{2}(2a\rho a^{\dagger}-a^{\dagger}a\rho-\rho a^{\dagger}a) \notag\\
     &&+\frac{\gamma_{m}}{2}(\bar{n}_{m}+1)(2b\rho b^{\dagger}-b^{\dagger}b\rho-\rho b^{\dagger}b) \notag \\
     &&+\frac{\gamma_{m}}{2}\bar{n}_{m}(2b^{\dagger}\rho b-bb^{\dagger}\rho-\rho bb^{\dagger}),
\end{eqnarray}
where we assume that the cavity field is connected with a vacuum bath. $\gamma_{m}$ represents the mechanical decay, and $\bar{n}_{m}$ is the average thermal photon number related to the temperature by $\bar{n}_{m}=[\text{exp}(\omega_{m}/k_{B}T_{M})-1]^{-1}$, where $k_{B}$ is the Boltzmann constant, $T_{M}$ is the temperature of the enviroment.

\subsection{\label{sec:level21}1PB and 2PB without mechanical pump}
\begin{figure}
  \centering
  \includegraphics[width=0.47\textwidth]{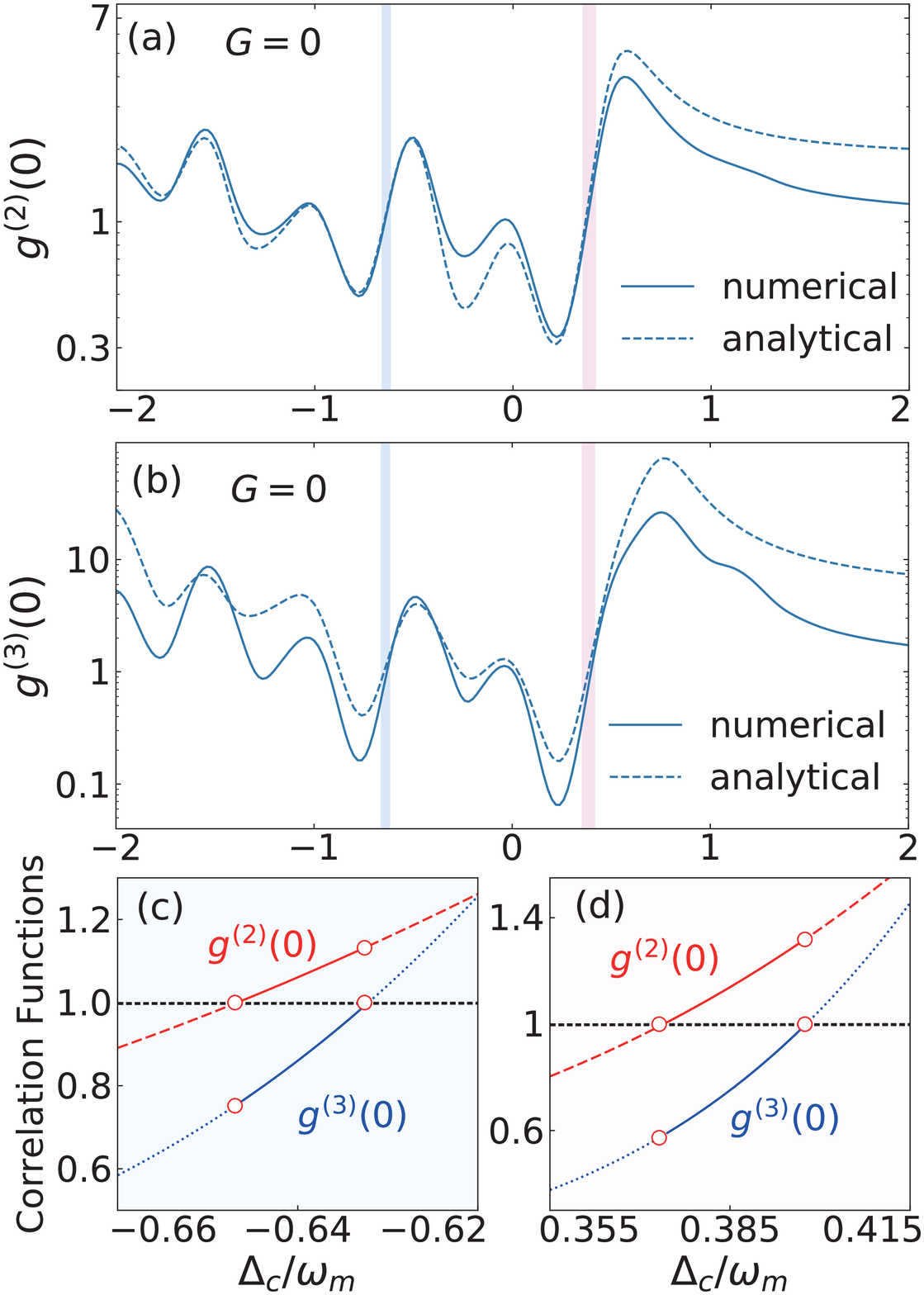}
  \caption{The correlation functions $g^{(2)}(0)$ and $g^{(3)}(0)$ versus $\Delta_{c}/\omega_{m}$ for the OMS without mechanical pump are show in (a),(b) repectively. The solid curves are based the numerical solution of Eq.~(\ref{master equation}), while the dashed curves are based on the analytical solution of Eq.~(\ref{longterm solution}). The case of $g^{(2)}(0)\ll1$ (i.e., the dip) in (a) indicates 1PB. (c),(d) Correlation functions $g^{(2)}(0)$ and $g^{(3)}(0)$ versus $\Delta_{c}/\omega_{m}$ without mechanical pump. The horizontal black dashed lines show $g^{(n)}(0)=1(n,m=2,3)$ on the basis of Eq.~(\ref{2PB}). The solid lines represent the parameter space satisfying the criterion, the blue dotted and red dashed lines represent the region that does not satisfy the criterion. Figure~\ref{fig3}(c) shows 2PB with the two-phonon sidebands, corresponding to $\vert 0\rangle _{a}\vert \tilde{0}(0)\rangle _{b}\leftrightarrow\vert 2\rangle _{a}\vert \tilde{0}(2)\rangle_{b}$. Figure~\ref{fig3}(d) shows 2PB without phonon sidebands, corresponding to $\vert 0\rangle _{a}\vert \tilde{0}(0)\rangle _{b}\leftrightarrow\vert 2\rangle_{a}\vert \tilde{2}(2)\rangle_{b}$. The parameters are taken as $g_{0}/\omega_{m}=0.5$, $\Omega/\omega_{m}=0.01$, $\gamma_{c}/\omega_{m}=0.3$, $\gamma_{m}/\omega_{m}=0.001$ and $\bar{n}_{m}=0$ ($T=0$).}   \label{fig2}
\end{figure}
In the OMS without the mechanical pump, i.e., $G=0$, we get the approximate solution of the equal-time second-order and third-order correlation functions
\begin{eqnarray}
  g^{(2)}(0) &=& \frac{4\chi_{1}^{2}+\gamma_{c}^{2}}{4\chi_{2}^{2}+\gamma_{c}^{2}}, \label{g2formula}\\
  g^{(3)}(0) &=& \frac{4(\chi_{1}^{2}+\gamma_{c}^{2})^{2}}{(4\chi_{2}^{2}+\gamma_{c}^{2})(4\chi_{3}^{2}+\gamma_{c}^{2})}, \label{g3formula}
\end{eqnarray}
where $\chi_{n}=\Delta_{c}-n\eta$. 

Supposing that the driving field is tuned to the single-photon resonantce (SPR) transition frequency, i.e., $\Delta_{c}=g_{0}^{2}/\omega_{m}$, the correlation function becomes
\begin{equation}\label{single correlation function}
 g_{\mathrm{SPR}}^{(2)}(0)=\frac{\gamma_{c}^{2}}{4\eta^{2}+\gamma_{c}^{2}}.
\end{equation}
In the strong-coupling regime, i.e., $g_{0}>\gamma_{c}$, we have $g_{\text{SPR}}^{(2)}(0)\ll1$. It means the probability of exciting the single-photon state is higher than that of preparing a two-photon state.%Single-photon nonlinearities in a strongly driven optomechanical system with quadratic coupling
%However, because both $\delta_{2}-2\delta_{1}=2g_{0}^{2}/\omega_{m}$ and $\gamma_{c}$ do not contain mechanical pump strength $G$, the whole formula has nothing to do with mechanical pump. So the mechanical pump does not impact the two order correlation function $ g_{\mathrm{SPR}}^{(2)}(0)$ at all in addition to changing the frequency of cavity resonance. The larger $g_{0}$ is, the small the correlation funtion $g_{SPR}^{(2)}(0)$ is.
%In the two-photon resonance (TPR) transition $\left|0\right\rangle _{a}\left|\tilde{0}(0)\right\rangle _{b}\rightarrow\left|2\right\rangle _{a}\left|\tilde{0}(2)\right\rangle _{b}$, the correlation function becomes

In the case of two-photon resonance (TPR), $\Delta_{c}=2g_{0}^{2}/\omega_{m}$,
the equal-time second-order correlation function becomes
\begin{equation}\label{two correlation function}
g_{\mathrm{TPR}}^{(2)}(0)=\frac{4\eta^{2}+\gamma_{c}^{2}}{\gamma_{c}^{2}}.
\end{equation}
%%%Single-photon nonlinearities in a strongly driven optomechanical system with quadratic coupling
We have $g^{(2)}(0) >\gg1$, which indicates that the cavity tends to be in the two-photon state rather than be the single-phonon state. %Single-photon nonlinearities in a strongly driven optomechanical system with quadratic coupling
The single-photon or two-phonon transitions can also happen in the $n$-phonon sidebands, as discussed later. %Phonon blockade in a quadratically coupled optomechanical system
%(b),(d)%in Figs. 4(a) and 4(b),

To study 1PB, we calculate the optical correlation function $g^{(2)}(0)$ by using both analytic and numerical method. The condition $g^{(2)}(0)\ll1$ characterizes 1PB. In order to prove 2PB where the absorption of two photons suppresses the absorption of further photons, it is sufficient to fulfill a necessary criterion, i.e.,
\begin{eqnarray}\label{2PB}
  g^{(2)}(0) &>& 1, \nonumber \\
  g^{(3)}(0) &<& 1.
\end{eqnarray}

In Figs.~\ref{fig2}(a) and \ref{fig2}(b), we plot both the optical correlation functions $g^{(2)}(0)$ and $g^{(3)}(0)$ versus $\Delta_{c}/\omega_{m}$ for $G=0$, of which the analytical and numerical results fit well. In general, $g^{(2)}(0)>1$ stands for a supper-Poisson distribution of the cavity field and $g^{(2)}(0)\gg1$ corresponds to photon-induced tunneling (PIT). $g^{(2)}(0)<1$ represents the sub-Poisson statistics and $g^{(2)}(0)\ll1$ corresponds to 1PB signifying nonclassical correlation. The condition $g^{(2)}(0)\rightarrow0$ means a complete 1PB. As shown in Fig.~\ref{fig2}(a), 1PB (i.e., the dip) occurs. The dip corresponds to 1PB and also the SPR case relating to the single-photon process $\vert0\rangle _{a}\vert \tilde{0}(0)\rangle _{b}\leftrightarrow\vert 1\rangle _{a}\vert \tilde{0}(1)\rangle _{b}$.
The peak corresponds to PIT and also the two-photon process $\vert 0\rangle _{a}\vert \tilde{0}(0)\rangle _{b}\leftrightarrow\vert 2\rangle _{a}\vert \tilde{0}(2)\rangle_{b}$.
As a matter of fact, the photon transitions can happen in the $n$-photon sidebands. Figures~\ref{fig2}(c) and \ref{fig2}(d) show the correlation functions $g^{(2)}(0)$ and $g^{(3)}(0)$ versus $\Delta_{c}/\omega_{m}$ without mechanical pump. We find 2PB emerges around $\Delta/\omega_{m}=-0.64$ or $\Delta/\omega_{m}=0.385$, corresponding to the transition $\vert0\rangle _{a}\vert \tilde{0}(0)\rangle _{b}\leftrightarrow\vert2\rangle_{a}\vert \tilde{0}(2)\rangle_{b}$, or two-phonon sideband $\vert0\rangle _{a}\vert \tilde{0}(0)\rangle_{b}\leftrightarrow\vert 2\rangle_{a}\vert \tilde{2}(2)\rangle _{b}$, respectively, which fulfills the correlation given in Eq.~(\ref{xi2}).
\subsection{\label{sec:level22}Mechanical engineering of 1PB and 2PB}
%Experimentally, a mechanical oscillator with an integrated electrical interface can be mechanically pumped by applying an radio frequency voltage~\cite{bowen2017} as shown in Fig.~\ref{fp}(b) and this device is identical to that studied in Ref.~\cite{38bowen2016}, which can be referred to for details on device fabrication. The amplitude $G$ can be a arbitrary value theoretically~\cite{bowen2017} by adjusting the bias gate voltages $V_{\text{DC}}$ and $V_{\text{AC}}$. The %weak%
%mechanical pumping field can also be realised by an dc voltage source~\cite{bowen2013} or a piezoelectric radio frequency signal~\cite{LFan2015}. Recently, this OMS with mechanical pump, based on eletro-optomechanical devices, has been realized in the experiment~\cite{bowen2017,bowen2013} as shown in Fig.~\ref{fp}(b).%Optically induced phonon blockade in an optomechanical system with second-order nonlinearity

In the OMS with the mechanical pump, the equal-time second-order correlation and third-order correlations are modified as
\begin{eqnarray}
g^{(2)}(0)&=& \frac{4(\chi_{1}-\delta)^{2}+\gamma_{c}^{2}}{4(\chi_{2}-\delta)^{2}+\gamma_{c}^{2}}, \label{g2_2}\\
g^{(3)}(0)&=&\frac{4[(\chi_{1}-\delta)^{2}+\gamma_{c}^{2}]^{2}}{[4(\chi_{2}-\delta)^{2}+\gamma_{c}^{2}][4(\chi_{3}-\delta)^{2}+\gamma_{c}^{2}]}. \label{g3_2}
\end{eqnarray}

For the SPR case, $\Delta_{c}=g_{0}^{2}+2g_{0}G/\omega_{m}$, the equal-time second-order correlation function is given as:
\begin{equation}\label{single correlation function}
 g_{\mathrm{SPR}}^{(2)}(0)=\frac{\gamma_{c}^{2}}{4\eta^{2}+\gamma_{c}^{2}}.
\end{equation}

For the TPR case, $\Delta_{c}=(2g_{0}^{2}+2g_{0}G)/\omega_{m}$, the equal-time second-order correlation function is given as:
\begin{equation}\label{two correlation function}
g_{\mathrm{TPR}}^{(2)}(0)=\frac{4\eta^{2}+\gamma_{c}^{2}}{\gamma_{c}^{2}}.
\end{equation}

As we can see, the mechanical pump only shifts the optical driving frequency of photon resonances, but doesn't weak the strength of equal-time correlation functions.
\begin{figure}
  \centering
  \includegraphics[width=0.47 \textwidth]{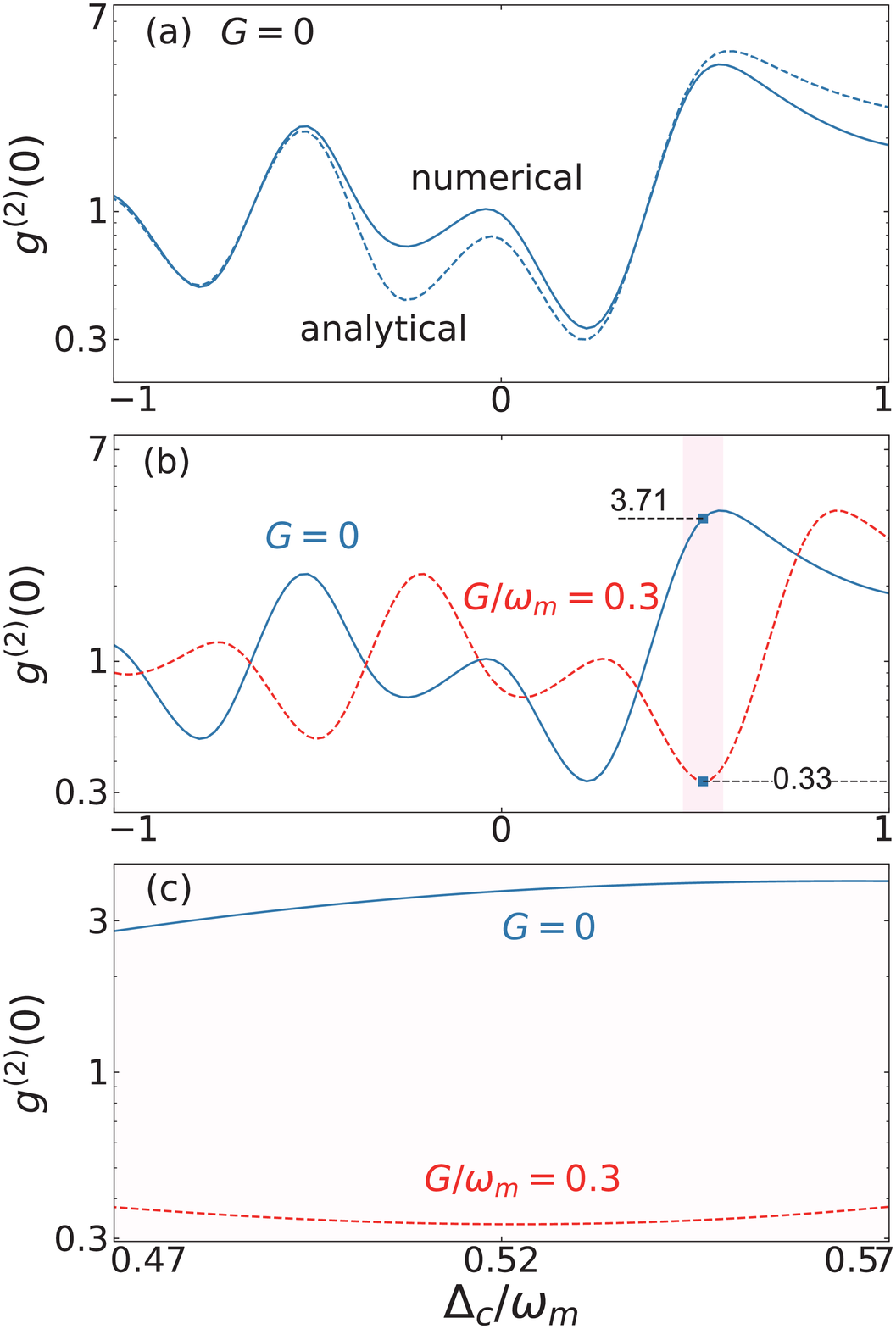}
  \caption{(a) The equal-time second-order correlation function $g^{(2)}(0)$ versus $\Delta_{c}/\omega_{m}$ for the system without mechanical pump, i.e., $G=0$. The dashed curve is the theoretically predicted second-order correlation function and the solid line is from numerical results. (b),(c) numerical equal-time second-order correlation function versus $\Delta_{c}/\omega_{m}$ for $G=0$ and $G/\omega_{m}=0.3$ respectively. The other parameters are the same as Fig.~\ref{fig2}.}\label{fig3}
\end{figure}

We now consider the mechanical engineering of $1$PB. In Fig.~\ref{fig3}(a), we plot both the analytical and numerical correlation function $g^{(2)}(0)$ versus $\Delta_{c}/\omega_{m}$ for $G/\omega_{m}=0.3$, and the analytical and numerical results are in good agreement. The dashed curve is based on the analytical solutions while the solid curve is based on the numerical results. For the OMS without mechanical pump, $ g^{(2)}(0)$ always has a dip at the specific optical detuning $\Delta_{c}/\omega_{m}=0.22\omega_{m}$ or a peak at $\Delta_{c}/\omega_{m}=0.55\omega_{m}$, corresponding to 1PB or PIT, respectively. In contrast, with mechanical pump, by tuning the mechanical strength, a shift for 1PB can be achieved as shown in Fig.~\ref{fig3}(b) and ~\ref{fig3}(c), i.e., $g^{(2)}(0)=3.71$ (no 1PB) for $G=0$, $g^{(2)}(0)=0.33$ (1PB) for $G/\omega_{m}=0.3$. We note that the shift of correlation functions is corresponding to the energy shift given in Eq~(\ref{xi2}) as mentioned above. The shift can also quantitatively derived by comparing Eq.~(\ref{g2formula}) and  Eq.~(\ref{g2_2}). This implies the mechanical engineering of a purely quantum effect, i.e., 1PB. Due to the mechanical strength is tunable, it is also possible to prerpare more feasible single-photon sources with multi-frequencies, which is fundamentally different from the previous studies.
\begin{figure}
  \centering
  \includegraphics[width=0.47\textwidth]{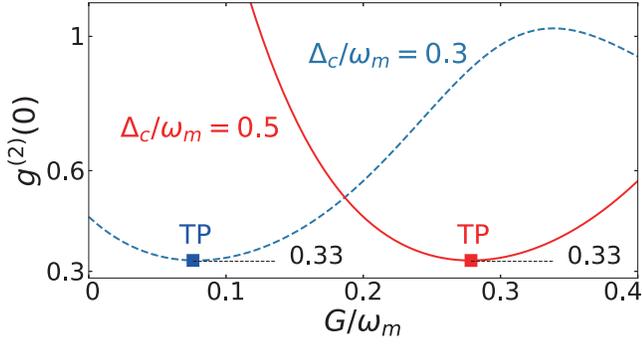}
  \caption{The equal-time second-order correlation function $g^{(2)}(0)$ versus the strength of mechanical pump $G$ for the conditions $\Delta_{c}/\omega_{m}=0.3$ and $\Delta_{c}/\omega_{m}=0.5$, respectively. The other parameters are the same as Fig.~\ref{fig2}. }   \label{g2G055_01}
\end{figure}

In Fig.~\ref{g2G055_01}, we numerically plot correlation function $g^{(2)}(0)$ versus the strength of mechanical pump $G$ under the conditions $\Delta_{c}/\omega_{m}=0.3$ and $\Delta_{c}/\omega_{m}=0.5$ respectively.
% The numerical result is shown in Fig.~\ref{g2G055_01}. %句式来自网上文献
Clearly for higher mechanical strength, the correlation function gradually changed due to the further shifted energy anharmonicity until reaching the lowest turning point (TP), where 1PB occurs. By deriving the correlation function of Eq.~(\ref{g2formula}), the mechanical strength at TP with the fixed detuning $\Delta_{c}$ then can be given:%使用的网上句式
\begin{equation}\label{g2formula2}
G=\frac{-3g_{0}^{2}+\sqrt{g_{0}^{4}+\gamma_{c}^{2}}+2\Delta_{c}\omega_{m}}{4g_{0}}.
\end{equation}
\begin{figure}
  \centering
  \includegraphics[width=0.47 \textwidth]{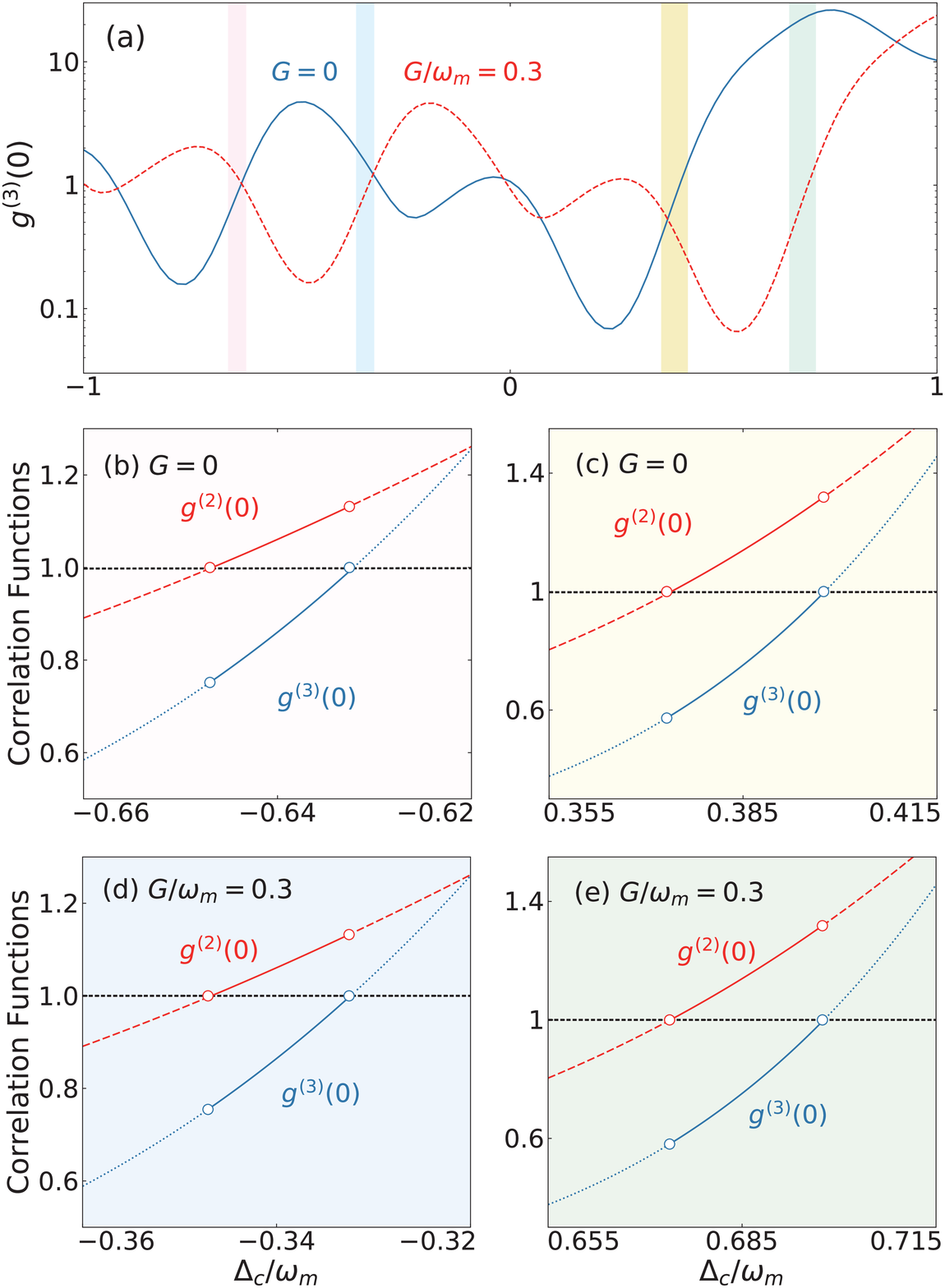}
  \caption{(a) The equal-time third-order correlation function $g^{(3)}(0)$ as a function of $\Delta_{c}/\omega_{m}$ for the strength of $G=0$ (blue solid line) and $G/\omega_{m}=0.5$ (red dashed line) respectively. (b),(c) Correlation functions $g^{(2)}(0)$ (red lines) and $g^{(3)}(0)$ (blue lines) versus $\Delta_{c}/\omega_{m}$ around $\Delta_{c}/\omega_{m}=-0.64$ and $\Delta_{c}/\omega_{m}=0.385$ respectively, with the mechanical strength of $G=0$. (d),(e) Correlation functions $g^{(2)}(0)$ (red lines) and $g^{(3)}(0)$ (blue lines) versus $\Delta_{c}/\omega_{m}$ around $\Delta_{c}/\omega_{m}=-0.34$ and $\Delta_{c}/\omega_{m}=0.685$ respectively, with the mechanical strength of $G/\omega_{m}=0.3$. The horizontal black dashed lines show $g^{(n)}(0)=1(n,m=2,3)$ on the basis of Eq.~(\ref{2PB}). The solid lines in (b-e) represent the parameter space satisfying the criterion, the blue dotted and red dashed lines represent the region that does not satisfy the criterion. The other parameters are the same as Fig.~\ref{fig2}.}\label{tprsub12_1}
\end{figure}

Owing to the fact that 1PB can be engineered by the mechanical pump, we now consider 2PB. In the following, we show 2PB and the mechanical engineering of 2PB can also be achieved with the mechanical pump.

%In order to prove two-photon blockade, it is sufficient for photon number correlation to fulfill the criterion
%%%
In Fig.~\ref{tprsub12_1}(a), we show equal-time third-order correlation function versus the driving detuning $\Delta_{c}/\omega_{m}$ from $G=0$ to $G/\omega_{m}=0.3$, which also be shifted owing to the mechanical pump. Figures~\ref{tprsub12_1}(b)-(e) show correlation functions $g^{(2)}(0)$ and $g^{(2)}(0)$ versus the driving detuning $\Delta_{c}/\omega_{m}$. The horizontal black dashed lines show $g^{(n)}(0)=1(n,m=2,3)$ on the basis of Eq.~(\ref{2PB}). The solid lines represent the parameters that satisfy the criterion; the blue dotted and red dashed lines represent the region that does not satisfy the criterion. We have found that there are two optical detunings where 2PB occurs. Figures~\ref{tprsub12_1}(b) and \ref{tprsub12_1}(d) indicate 2PB with the two-phonon sidebands, corresponding to $\vert 0\rangle _{a}\vert \tilde{0}(0)\rangle _{b}\leftrightarrow\vert 2\rangle _{a}\vert \tilde{0}(2)\rangle_{b}$. Figures~\ref{tprsub12_1}(c) and ~\ref{tprsub12_1}(e) indicate 2PB without phonon sidebands, corresponding to $\vert 0\rangle _{a}\vert \tilde{0}(0)\rangle _{b}\leftrightarrow\vert 2\rangle _{a}\vert \tilde{2}(2)\rangle_{b}$. Obviously, when we change the strength of the mechanical pump from $G=0$ to $G/\omega_{m}=0.3$, 2PB occurs from $\Delta_{c}/\omega_{m}=-0.64$ to $\Delta_{c}/\omega_{m}=-0.34$ with two phonon sidebands, and from $\Delta_{c}/\omega_{m}=0.385$ to $\Delta_{c}/\omega_{m}=0.685$ without phonon sidebands. We note that the difference between two detunings is $\Delta_{c}/\omega_{m}=0.3$ exactly as predicted from fomula (\ref{g3formula})~and fomula~(\ref{g3_2}).
%The lines represent the shift in the NMR shielding value句式

\begin{figure}
  \centering
  \includegraphics[width=0.47 \textwidth]{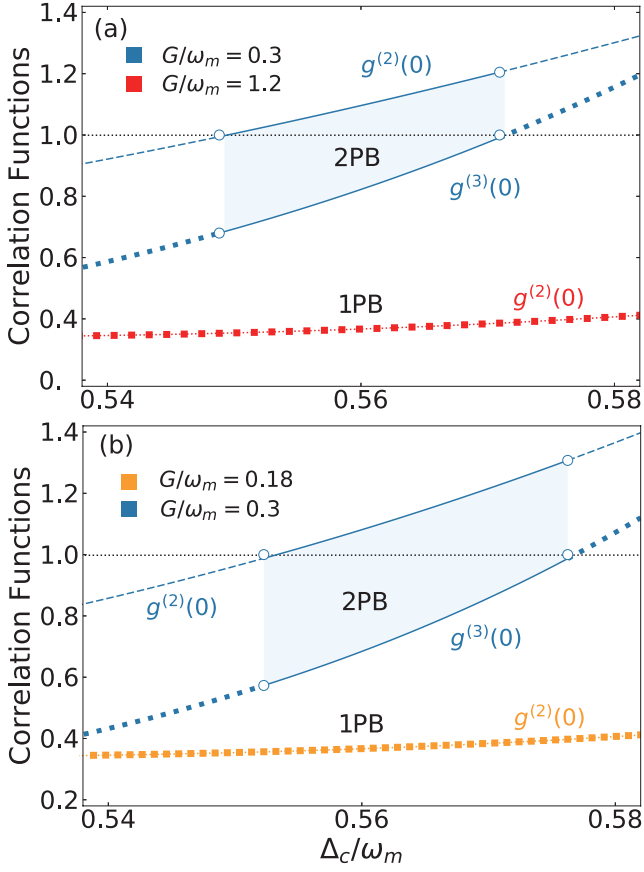}
  \caption{(a),(b) The correlation functions $g^{(2)}(0)$ and $g^{(3)}(0)$ versus $\Delta_{c}/\omega_{m}$. The color codings identify the different strengths of mechanical pump: $G/\omega_{m}=0.3$ (blue lines), $G/\omega_{m}=1.2$ (red line), and $G/\omega_{m}=0.18$ (orange line). The blue backgrounds are regions where 1PB and 2PB can occur at the same driving detuning, with two different strengths of mechanical pump respectively. The other parameters are the same as Fig.~\ref{fig2}.}\label{Fig6}
\end{figure}
%%The color coding identify the two interesting phenomena that may happen: 1PB (blue), 2PB (magenta)
%%%

Figure~\ref{Fig6} shows the correlation functions $g^{(2)}(0)$ and $g^{(3)}(0)$ versus $\Delta_{c}/\omega_{m}$. The color codings identify the different strengths of mechanical pump: $G/\omega_{m}=0.3$ (blue lines), $G/\omega_{m}=1.2$ (red line), and $G/\omega_{m}=0.18$ (orange line). The blue backgrounds are regions where 1PB and 2PB can occur at the same driving detuning, with two different strengths of mechanical pump respectively. In the presence of mechanical pump, we can achieve both 1PB and 2PB using the same driving laser as indicated by Figs.~\ref{Fig6}(a) and \ref{Fig6}(b). Figure~\ref{Fig6}(a) corresponds to 2PB without phonon sideband and Fig.~\ref{Fig6}(b) corresponds to 2PB with two-phonon sidebands. The other parameters are the same as Fig.~\ref{fig2}.
\section{CONCLUSION AND OUTLOOK}\label{OUTLOOKsec}
In this paper, we analytically and numerically calculate the equal-time second-order and third-order correlation functions of a cavity OMS with a mechanical pump. By properly choosing the strength of mechanical pump, we find the following: (i) selective 1PB or 2PB can be achieved at the adjustable optical detuning. (ii) More interestingly, simultaneous 1PB and 2PB can be achieved at the
same driving frequency. This indicates that the mechanical engineering of OMS can provide more flexile control about few-photon emissions. Our work shows that OMS can become another promising platform to achieve such a goal. We note that in very recent experiments, PB or single-photon emission was also observed in driven pendulum-resonator system~\cite{G. S. Paraoanu2019} or acoustically-driven quantum well system~\cite{T.-K. Hsiao2019}.

Our work can be further extended to study mechanical engineering of more purely quantum optomechanical effects, such as mechanical squeezing, photon-phonon entanglement. Moreover, due to the mechanical pump's exceptional operability and convenience nature, the OMS with the mechanical pump is an excellent candidate for exploring new applications from precision metrology to tunable photonics. More interesting than direct-current (scalar) pump, the alternating current mechanical pump in the OMS, including phase effects, will be studied in future work.

\textit{Note added}. After finishing this work, we became aware of a related work about mechanically controlled single-photon emitter and frequncy comb, by using a membrane-spin hybrid device~\cite{Plenio2019}.

%The previous paper shows that strong coupling of the micromechanical oscillator to an optical cavity will result in nonlinear phenomenon like 1PB~\cite{Prabl}. In this article, we give the eigenvalue eigenstates of the system and study PB in a linear mechanical pumped optomechanical cavity. Compared with the PB in OMS without mechanical pump, we find that the OMS in presence pump can mechanical control the PBs by adjusting the mechanical pump strength. We By analytically and numerically analyzing the correlation functions of the system, we can realize the mechanical engineering of PB, that is, to adjust the appropriate mechanical pump strength, we can achieve PB in any frequency of optical driving.

\section{ACKNOWLEDGMENTS}
We thank HuiLai Zhang and Tao Liu for useful discussions. This work is supported by NSF of China under Grants No. 11474087 and No. 11774086, and the
HuNU Program for Talented Youth.

\appendix
\begin{widetext}

\section{Expansion for a unitary operator $D$ up to the order 1}

We consider the limit case of $g_{0}/\omega_{m}\ll1$. In this case,
we can expand the displacement operators $\text{exp}[g_{0}/\omega_{m}(b-b^{\dagger})]$
to $1-g_{0}/\omega_{m}(b-b^{\dagger})$:

\begin{eqnarray}
\left.\left\langle \tilde{m}^{\prime}(1)\right|\tilde{0}(0)\right\rangle _{b} & = & \delta_{m^{\prime},0}+\frac{g_{0}}{\omega_{m}}\delta_{m^{\prime},1},
\end{eqnarray}
\begin{equation}
\left.\left\langle \tilde{m}(1)\right|\tilde{0}\right\rangle _{b}=\delta_{m,0}+\frac{g_{0}}{\omega_{m}}\delta_{m,1},
\label{m 10}
\end{equation}
\begin{equation}\label{eq:q10}
\left.\left\langle \tilde{q}(1)\right|\tilde{0}\right\rangle _{b}=\delta_{q,0}+\frac{g_{0}}{\omega_{m}}\delta_{q,1},
\end{equation}
\begin{equation}\label{q1 0}
\left.\left\langle \tilde{m}(2)\right|\tilde{m}^{\prime}(1)\right\rangle _{b}=\delta_{m,m^{\prime}}+\sqrt{m^{\prime}+1}\frac{g_{0}}{\omega_{m}}\delta_{m,m^{\prime}+1}-\sqrt{m^{\prime}}\frac{g_{0}}{\omega_{m}}\delta_{m,m^{\prime}-1},
\end{equation}
\begin{equation}\label{m2 q1}
\left.\left\langle \tilde{m}^{\prime}(2)\right|\tilde{q}(1)\right\rangle _{b}=\delta_{m^{\prime},q}+\sqrt{q+1}\frac{g_{0}}{\omega_{m}}\delta_{m^{\prime},q+1}-\sqrt{q}\frac{g_{0}}{\omega_{m}}\delta_{m^{\prime},q-1},
\end{equation}
\begin{equation}\label{m3 m2}
\left.\left\langle \tilde{m}(3)\right|\tilde{m}^{\prime}(2)\right\rangle _{b}=\delta_{m,m^{\prime}}+\sqrt{m^{\prime}+1}\frac{g_{0}}{\omega_{m}}\delta_{m,m^{\prime}+1}-\sqrt{m^{\prime}}\frac{g_{0}}{\omega_{m}}\delta_{m,m^{\prime}-1}.
\end{equation}

\section{Equal-time third-order correlation function}
For the correlations given in Eq.~(\ref{m 10}), we get the single-photon
probability:
\begin{align}
P_{1} & =\sum_{m=0}^{\infty}\left|C_{1}\right|^{2}\nonumber \\
 & =\sum_{m=0}^{\infty}\left|\frac{-\Omega\left.\left\langle \tilde{m}(1)\right|\tilde{0}\right\rangle _{b}e^{-iE_{0,0}t}}{\Delta_{c}+m\omega_{m}-\frac{g_{0}^{2}+2g_{0}G}{\omega_{m}}-i\frac{\gamma_{c}}{2}}\right|^{2}\nonumber \\
 & =\sum_{m=0}^{\infty}\frac{\Omega^{2}\left|\left.\left\langle \tilde{m}(1)\right|\tilde{0}\right\rangle _{b}\right|^{2}}{\left(\Delta_{c}+m\omega_{m}-\frac{g_{0}^{2}+2g_{0}G}{\omega_{m}}\right)^{2}+\left(\frac{\gamma_{c}}{2}\right)^{2}}\nonumber \\
 & =\frac{\Omega^{2}}{\left(\Delta_{c}-\frac{g_{0}^{2}+2g_{0}G}{\omega_{m}}\right)^{2}+\left(\frac{\gamma_{c}}{2}\right)^{2}}+\frac{\Omega^{2}\beta^{2}}{\left(\Delta_{c}+\omega_{m}-\frac{g_{0}^{2}+2g_{0}G}{\omega_{m}}\right)^{2}+\left(\frac{\gamma_{c}}{2}\right)^{2}}.\label{eq:P1lost}
\end{align}
For three-photon state,
\begin{eqnarray}
C_{3,m}(t) & = & -\sqrt{6}\Omega^{3}\sum_{m^{\prime},q=0}^{\infty}\frac{\left.\left\langle \tilde{m}(3)\right|\tilde{m}^{\prime}(2)\right\rangle _{b}\left.\left\langle \tilde{m}^{\prime}(2)\right|\tilde{q}(1)\right\rangle _{b}\left.\left\langle \tilde{q}(1)\right|\tilde{0}\right\rangle _{b}e^{-iE_{0,0}t}}{\left(E_{1,q}-E_{0,0}-i\frac{\gamma_{c}}{2}\right)\left(E_{2,m^{\prime}}-E_{0,0}-i\gamma_{c}\right)\left(E_{3,m}-E_{0,0}-i\frac{3\gamma_{c}}{2}\right)},
\end{eqnarray}
and three-photon state probability reads
\begin{eqnarray}
P_{3} & = & \sum_{m=0}^{\infty}\left|C_{3}\right|^{2}\nonumber \\
 &  & \sum_{m=0}^{\infty}\left|\sum_{m^{\prime},q=0}^{\infty}\frac{-\sqrt{6}\Omega^{3}\left.\left\langle \tilde{m}(3)\right|\tilde{m}^{\prime}(2)\right\rangle _{b}\left.\left\langle \tilde{m}^{\prime}(2)\right|\tilde{q}(1)\right\rangle _{b}\left.\left\langle \tilde{q}(1)\right|\tilde{0}\right\rangle _{b}e^{-iE_{0,0}t}}{\left(E_{1,q}-E_{0,0}-i\frac{\gamma_{c}}{2}\right)\left(E_{2,m^{\prime}}-E_{0,0}-i\gamma_{c}\right)\left(E_{3,m}-E_{0,0}-i\frac{3\gamma_{c}}{2}\right)}\right|^{2}\nonumber \\
 & = & 6\Omega^{6}\sum_{m=0}^{\infty}\left|\sum_{m^{\prime}=0}^{\infty}\left.\left\langle \tilde{m}(3)\right|\tilde{m}^{\prime}(2)\right\rangle _{b}U\right|^{2}\nonumber \\
 & = & 6\Omega^{6}\sum_{m=0}^{\infty}\left|V\right|^{2},
\end{eqnarray}
where
\begin{equation}\label{eq:U}
U=\sum_{q=0}^{\infty}\frac{\left.\left\langle \tilde{m}^{\prime}(2)\right|\tilde{q}(1)\right\rangle _{b}\left.\left\langle \tilde{q}(1)\right|\tilde{0}\right\rangle _{b}}{\left(E_{1,q}-E_{0,0}-i\frac{\gamma_{c}}{2}\right)\left(E_{2,m^{\prime}}-E_{0,0}-i\gamma_{c}\right)\left(E_{3,m}-E_{0,0}-i\frac{3\gamma_{c}}{2}\right)},
\end{equation}
and
\begin{equation}
V=\sum_{m^{\prime}=0}^{\infty}\left.\left\langle \tilde{m}(3)\right|\tilde{m}^{\prime}(2)\right\rangle _{b}U.\label{eq:V}
\end{equation}
Substitutiting Eqs.~(\ref{q1 0}) and (\ref{m2 q1}) into Eq.~(\ref{eq:U}),
we can obtain
\begin{eqnarray}\label{eq:U2}
U & = & \sum_{q=0}^{\infty}\frac{\left.\left\langle \tilde{m}^{\prime}(2)\right|\tilde{q}(1)\right\rangle _{b}\left.\left\langle \tilde{q}(1)\right|\tilde{0}\right\rangle _{b}}{\left(E_{1,q}-E_{0,0}-i\frac{\gamma_{c}}{2}\right)\left(E_{2,m^{\prime}}-E_{0,0}-i\gamma_{c}\right)\left(E_{3,m}-E_{0,0}-i\frac{3\gamma_{c}}{2}\right)}\nonumber \\
 & = & \sum_{q=0}^{\infty}\frac{\left(\delta_{m^{\prime},q}+\sqrt{q+1}\frac{g_{0}}{\omega_{m}}\delta_{m^{\prime},q+1}-\sqrt{q}\frac{g_{0}}{\omega_{m}}\delta_{m^{\prime},q-1}\right)\left(\delta_{q,0}+\frac{g_{0}}{\omega_{m}}\delta_{q,1}\right)}{\left(E_{1,q}-E_{0,0}-i\frac{\gamma_{c}}{2}\right)\left(E_{2,m^{\prime}}-E_{0,0}-i\gamma_{c}\right)\left(E_{3,m}-E_{0,0}-i\frac{3\gamma_{c}}{2}\right)}\nonumber \\
 & = & \frac{\delta_{m^{\prime},0}+\frac{g_{0}}{\omega_{m}}\delta_{m^{\prime},1}}{\left(E_{1,0}-E_{0,0}-i\frac{\gamma_{c}}{2}\right)\left(E_{2,m^{\prime}}-E_{0,0}-i\gamma_{c}\right)\left(E_{3,m}-E_{0,0}-i\frac{3\gamma_{c}}{2}\right)}\nonumber \\
 &  & +\frac{\frac{g_{0}}{\omega_{m}}\left(\delta_{m^{\prime},1}+\sqrt{2}\frac{g_{0}}{\omega_{m}}\delta_{m^{\prime},2}-\frac{g_{0}}{\omega_{m}}\delta_{m^{\prime},0}\right)}{\left(E_{1,1}-E_{0,0}-i\frac{\gamma_{c}}{2}\right)\left(E_{2,m^{\prime}}-E_{0,0}-i\gamma_{c}\right)\left(E_{3,m}-E_{0,0}-i\frac{3\gamma_{c}}{2}\right)}.
\end{eqnarray}
Substituting Eqs.~(\ref{m3 m2}) and (\ref{eq:U2}) into Eq.~(\ref{eq:V}),
we can obtain
\begin{eqnarray}
V & = & \sum_{m^{\prime}=0}^{\infty}\left.\left\langle \tilde{m}(3)\right|\tilde{m}^{\prime}(2)\right\rangle _{b}U\nonumber \\
 & = & \sum_{m^{\prime}=0}^{\infty}\frac{\left.\left\langle \tilde{m}(3)\right|\tilde{m}^{\prime}(2)\right\rangle _{b}\left(\delta_{m^{\prime},0}+\frac{g_{0}}{\omega_{m}}\delta_{m^{\prime},1}\right)}{\left(E_{1,0}-E_{0,0}-i\frac{\gamma_{c}}{2}\right)\left(E_{2,m^{\prime}}-E_{0,0}-i\gamma_{c}\right)\left(E_{3,m}-E_{0,0}-i\frac{3\gamma_{c}}{2}\right)}\nonumber \\
 &  & +\sum_{m^{\prime}=0}^{\infty}\frac{\frac{g_{0}}{\omega_{m}}\left.\left\langle \tilde{m}(3)\right|\tilde{m}^{\prime}(2)\right\rangle _{b}\left(\delta_{m^{\prime},1}+\sqrt{2}\frac{g_{0}}{\omega_{m}}\delta_{m^{\prime},2}-\frac{g_{0}}{\omega_{m}}\delta_{m^{\prime},0}\right)}{\left(E_{1,0}-E_{0,0}-i\frac{\gamma_{c}}{2}\right)\left(E_{2,m^{\prime}}-E_{0,0}-i\gamma_{c}\right)\left(E_{3,m}-E_{0,0}-i\frac{3\gamma_{c}}{2}\right)}\nonumber \\
 & = & \sum_{m^{\prime}=0}^{\infty}\frac{\left(\delta_{m,m^{\prime}}+\sqrt{m^{\prime}+1}\frac{g_{0}}{\omega_{m}}\delta_{m,m^{\prime}+1}-\sqrt{m^{\prime}}\frac{g_{0}}{\omega_{m}}\delta_{m,m^{\prime}-1}\right)\left(\delta_{m^{\prime},0}+\frac{g_{0}}{\omega_{m}}\delta_{m^{\prime},1}\right)}{\left(E_{1,0}-E_{0,0}-i\frac{\gamma_{c}}{2}\right)\left(E_{2,m^{\prime}}-E_{0,0}-i\gamma_{c}\right)\left(E_{3,m}-E_{0,0}-i\frac{3\gamma_{c}}{2}\right)}\nonumber \\
 &  & +\sum_{m^{\prime}=0}^{\infty}\frac{\frac{g_{0}}{\omega_{m}}\left(\delta_{m,m^{\prime}}+\sqrt{m^{\prime}+1}\frac{g_{0}}{\omega_{m}}\delta_{m,m^{\prime}+1}-\sqrt{m^{\prime}}\frac{g_{0}}{\omega_{m}}\delta_{m,m^{\prime}-1}\right)\left(\delta_{m^{\prime},1}+\sqrt{2}\frac{g_{0}}{\omega_{m}}\delta_{m^{\prime},2}-\frac{g_{0}}{\omega_{m}}\delta_{m^{\prime},0}\right)}{\left(E_{1,1}-E_{0,0}-i\frac{\gamma_{c}}{2}\right)\left(E_{2,m^{\prime}}-E_{0,0}-i\gamma_{c}\right)\left(E_{3,m}-E_{0,0}-i\frac{3\gamma_{c}}{2}\right)}.
\end{eqnarray}
When $m^{\prime}=0$, we can obtain
\begin{eqnarray}
V_{0} & = & \frac{\delta_{m,0}+\frac{g_{0}}{\omega_{m}}\delta_{m,1}}{\left(E_{1,0}-E_{0,0}-i\frac{\gamma_{c}}{2}\right)\left(E_{2,0}-E_{0,0}-i\gamma_{c}\right)\left(E_{3,m}-E_{0,0}-i\frac{3\gamma_{c}}{2}\right)}\nonumber \\
 &  & +\frac{-\frac{g_{0}^{2}}{\omega_{m}^{2}}\left(\delta_{m,0}+\frac{g_{0}}{\omega_{m}}\delta_{m,1}\right)}{\left(E_{1,1}-E_{0,0}-i\frac{\gamma_{c}}{2}\right)\left(E_{2,0}-E_{0,0}-i\gamma_{c}\right)\left(E_{3,m}-E_{0,0}-i\frac{3\gamma_{c}}{2}\right)}.
\end{eqnarray}
When $m^{\prime}=1$, we can obtain
\begin{eqnarray}
V_{1} & = & \frac{\frac{g_{0}}{\omega_{m}}\left(\delta_{m,1}+\sqrt{2}\frac{g_{0}}{\omega_{m}}\delta_{m,2}-\frac{g_{0}}{\omega_{m}}\delta_{m,0}\right)}{\left(E_{1,0}-E_{0,0}-i\frac{\gamma_{c}}{2}\right)\left(E_{2,1}-E_{0,0}-i\gamma_{c}\right)\left(E_{3,m}-E_{0,0}-i\frac{3\gamma_{c}}{2}\right)}\nonumber \\
 &  & +\frac{\frac{g_{0}}{\omega_{m}}\left(\delta_{m,1}+\sqrt{2}\frac{g_{0}}{\omega_{m}}\delta_{m,2}-\frac{g_{0}}{\omega_{m}}\delta_{m,0}\right)}{\left(E_{1,1}-E_{0,0}-i\frac{\gamma_{c}}{2}\right)\left(E_{2,1}-E_{0,0}-i\gamma_{c}\right)\left(E_{3,m}-E_{0,0}-i\frac{3\gamma_{c}}{2}\right)}.
\end{eqnarray}
When $m^{\prime}=2$, we can obtain
\begin{eqnarray}
V_{2} & = & \frac{\sqrt{2}\frac{g_{0}^{2}}{\omega_{m}^{2}}\left(\delta_{m,2}+\sqrt{3}\frac{g_{0}}{\omega_{m}}\delta_{m,3}-\sqrt{2}\frac{g_{0}}{\omega_{m}}\delta_{m,1}\right)}{\left(E_{1,1}-E_{0,0}-i\frac{\gamma_{c}}{2}\right)\left(E_{2,2}-E_{0,0}-i\gamma_{c}\right)\left(E_{3,m}-E_{0,0}-i\frac{3\gamma_{c}}{2}\right)}.
\end{eqnarray}
So we obtain
\begin{eqnarray}
V & = & V_{0}+V_{1}+V_{2}\nonumber \\
 & = & \frac{\delta_{m,0}+\frac{g_{0}}{\omega_{m}}\delta_{m,1}}{EHQ_{m}}+\frac{-\frac{g_{0}^{2}}{\omega_{m}^{2}}\left(\delta_{m,0}+\frac{g_{0}}{\omega_{m}}\delta_{m,1}\right)}{FHQ_{m}}\nonumber \\
 &  & +\frac{\frac{g_{0}}{\omega_{m}}\left(\delta_{m,1}+\sqrt{2}\frac{g_{0}}{\omega_{m}}\delta_{m,2}-\frac{g_{0}}{\omega_{m}}\delta_{m,0}\right)}{EIQ_{m}}\nonumber \\
 &  & +\frac{\frac{g_{0}}{\omega_{m}}\left(\delta_{m,1}+\sqrt{2}\frac{g_{0}}{\omega_{m}}\delta_{m,2}-\frac{g_{0}}{\omega_{m}}\delta_{m,0}\right)}{FIQ_{m}}\nonumber \\
 &  & +\frac{\sqrt{2}\frac{g_{0}^{2}}{\omega_{m}^{2}}\left(\delta_{m,2}+\sqrt{3}\frac{g_{0}}{\omega_{m}}\delta_{m,3}-\sqrt{2}\frac{g_{0}}{\omega_{m}}\delta_{m,1}\right)}{FJQ_{m}},
\end{eqnarray}
where
\begin{equation}
E=E_{1,0}-E_{0,0}-i\frac{\gamma_{c}}{2},
\end{equation}
\begin{equation}
F=E_{1,1}-E_{0,0}-i\frac{\gamma_{c}}{2},
\end{equation}
\begin{equation}
H=E_{2,0}-E_{0,0}-i\gamma_{c},
\end{equation}
\begin{equation}
I=E_{2,1}-E_{0,0}-i\gamma_{c},
\end{equation}
\begin{equation}
J=E_{2,2}-E_{0,0}-i\gamma_{c},
\end{equation}
\begin{equation}
Q_{m}=E_{3,m}-E_{0,0}-i\frac{3\gamma_{c}}{2}.
\end{equation}
When $m=0$, we obtain
\begin{eqnarray}
P_{3,m=0} & = & 6\Omega^{6}\sum_{m=0}\left|V\right|^{2}\nonumber \\
 & = & 6\Omega^{6}\sum_{m=0}^{\infty}\left|\frac{1}{EQ_{0}H}-\frac{g_{0}^{2}}{\omega_{m}^{2}FQ_{0}H}-\frac{g_{0}^{2}}{\omega_{m}^{2}EQ_{0}I}-\frac{g_{0}^{2}}{\omega_{m}^{2}FQ_{0}I}\right|^{2}\nonumber \\
 & = & 6\Omega^{6}\left|\frac{FI-\frac{g_{0}^{2}}{\omega_{m}^{2}}EI-\frac{g_{0}^{2}}{\omega_{m}^{2}}HF-\frac{g_{0}^{2}}{\omega_{m}^{2}}HE}{EFQ_{0}HI}\right|^{2}\nonumber \\
 & \approx & 6\Omega^{6}\left|\frac{1}{EQ_{0}H}\right|^{2}.
\end{eqnarray}
When $m=1$, we obtain
\begin{eqnarray}
P_{3,m=1} & = & 6\Omega^{6}\sum_{m=1}\left|V\right|^{2}\nonumber \\
 & = & 6\Omega^{6}\left|\frac{\frac{g_{0}}{\omega_{m}}}{EQ_{1}H}-\frac{\frac{g_{0}^{3}}{\omega_{m}^{3}}}{FQ_{1}H}+\frac{\frac{g_{0}}{\omega_{m}}}{EQ_{1}I}+\frac{\frac{g_{0}}{\omega_{m}}}{FQ_{1}I}-\frac{2\frac{g_{0}^{3}}{\omega_{m}^{3}}}{FQ_{1}J}\right|^{2}\nonumber \\
 & = & 6\Omega^{6}\left|\frac{\frac{g_{0}}{\omega_{m}}F-\frac{g_{0}^{3}}{\omega_{m}^{3}}E}{EFQ_{1}H}+\frac{\frac{g_{0}}{\omega_{m}}F+\frac{g_{0}}{\omega_{m}}E}{EFQ_{1}I}-\frac{2\frac{g_{0}^{3}}{\omega_{m}^{3}}}{FQ_{1}J}\right|^{2}\nonumber \\
 & \approx & 6\Omega^{6}\frac{g_{0}^{2}}{\omega_{m}^{2}}\left|\frac{1}{EQ_{1}H}+\frac{1}{EQ_{1}I}+\frac{1}{FQ_{1}I}\right|^{2}.
\end{eqnarray}
When $m=2$, we obtain
\begin{eqnarray}
P_{3,m=2} & = & 6\Omega^{6}\sum_{m=2}\left|V\right|^{2}\nonumber \\
 & = & 6\Omega^{6}\left|\frac{\sqrt{2}\frac{g_{0}^{2}}{\omega_{m}^{2}}}{EQ_{2}J}+\frac{\sqrt{2}\frac{g_{0}^{2}}{\omega_{m}^{2}}}{FQ_{2}J}+\frac{\sqrt{2}\frac{g_{0}^{2}}{\omega_{m}^{2}}}{FQ_{2}I}\right|^{2}\nonumber \\
 & = & 12\Omega^{6}\frac{g_{0}^{4}}{\omega_{m}^{4}}\left|\frac{1}{EQ_{2}J}+\frac{1}{FQ_{2}J}+\frac{1}{FQ_{2}I}\right|^{2}.
\end{eqnarray}
When $m=3$, we obtain
\begin{eqnarray}
P_{3,m=3} & = & 6\Omega^{6}\sum_{m=3}\left|V\right|^{2}\nonumber \\
 & = & 6\Omega^{6}\left|\frac{\sqrt{6}\frac{g_{0}^{3}}{\omega_{m}^{3}}}{FQ_{3}J}\right|^{2}\nonumber \\
 & = & 36\Omega^{6}\frac{g_{0}^{6}}{\omega_{m}^{6}}\left|\frac{1}{FQ_{3}J}\right|^{2}.
\end{eqnarray}

In the case of $\frac{g_{0}}{\omega_{m}}\ll1$ , the terms with high-order
can be safely neglected. Consequently the probabilities of finding
single, three photons in the cavity are, respectively, rewriten as:
\begin{equation}
P_{1}=\frac{\Omega^{2}}{\left(\Delta_{c}-\frac{g_{0}^{2}+2g_{0}G}{\omega_{m}}\right){}^{2}+\left(\frac{\gamma_{c}}{2}\right){}^{2}},\label{eq:getp1}
\end{equation}

\begin{equation}
P_{3}=6\Omega^{6}\left|\frac{1}{\left(E_{1,0}-E_{0,0}-i\frac{\gamma_{c}}{2}\right)\left(E_{2,0}-E_{0,0}-i\gamma_{c}\right)\left(E_{3,0}-E_{0,0}-i\frac{3\gamma_{c}}{2}\right)}\right|^{2}.\label{eq:getp3}
\end{equation}
For the eigenvalus given in Eq.~(\ref{xi2}) , we obtain
\begin{equation}
E_{0,0}=-\frac{G^{2}}{\omega_{m}},
\end{equation}

\begin{equation}
E_{1,0}=\Delta_{c}-\frac{g_{0}^{2}}{\omega_{m}}-\frac{2Gg_{0}}{\omega_{m}}-\frac{G^{2}}{\omega_{m}},
\end{equation}

\begin{equation}
E_{2,0}=2\Delta_{c}-\frac{4g_{0}^{2}}{\omega_{m}}-\frac{4Gg_{0}}{\omega_{m}}-\frac{G^{2}}{\omega_{m}},
\end{equation}

\begin{equation}
E_{3,0}=3\Delta_{c}-\frac{9g_{0}^{2}}{\omega_{m}}-\frac{6Gg_{0}}{\omega_{m}}-\frac{G^{2}}{\omega_{m}}.
\end{equation}
Substituing Eq. (\ref{eq:getp1}) and (\ref{eq:getp3}) into $g^{(3)}(0)=\frac{6P_{3}}{P_{1}^{3}}$
,we can obtain the equal-time third-order correlation
\begin{align}
g^{(3)}(0) & =\frac{6P_{3}}{P_{1}^{3}}\nonumber \\
 & =\frac{36\left[\left(\Delta_{c}-\frac{g_{0}^{2}+2g_{0}G}{\omega_{m}}\right)^{2}+\left(\frac{\gamma_{c}}{2}\right)^{2}\right]}{\left(E_{1,0}-E_{0,0}-i\frac{\gamma_{c}}{2}\right)^{2}\left(E_{2,0}-E_{0,0}-i\gamma_{c}\right)^{2}\left(E_{3,0}-E_{0,0}-i\frac{3\gamma_{c}}{2}\right)^{2}}\nonumber \\
 & =\frac{36\left[\left(\Delta_{c}-\frac{g_{0}^{2}+2g_{0}G}{\omega_{m}}\right)^{2}+\left(\frac{\gamma_{c}}{2}\right)^{2}\right]^{3}}{\left(\Delta_{c}-\frac{g_{0}^{2}+2Gg_{0}}{\omega_{m}}-i\frac{\gamma_{c}}{2}\right)^{2}\left(2\Delta_{c}-\frac{4g_{0}^{2}+4Gg_{0}}{\omega_{m}}-i\gamma_{c}\right)^{2}\left(3\Delta_{c}-\frac{9g_{0}^{2}+6Gg_{0}}{\omega_{m}}-i\frac{3\gamma_{c}}{2}\right)^{2}}\nonumber \\
 & =\frac{36\left[\left(\Delta_{c}-\eta-\delta\right)^{2}+\left(\frac{\gamma_{c}}{2}\right)^{2}\right]^{2}}{\left[4\left(\Delta_{c}-2\eta-\delta\right)^{2}+\gamma_{c}^{2}\right]\left[9\left(\Delta_{c}-3\eta-\delta\right)^{2}+\left(\frac{3\gamma_{c}}{2}\right)^{2}\right]}\nonumber \\
 & =\frac{36\left[4\left(\Delta_{c}-\eta-\delta\right)^{2}+4\left(\frac{\gamma_{c}}{2}\right)^{2}\right]^{2}}{4\left[4\left(\Delta_{c}-2\eta-\delta\right)^{2}+\gamma_{c}^{2}\right]\left[36\left(\Delta_{c}-3\eta-\delta\right)^{2}+4\left(\frac{3\gamma_{c}}{2}\right)^{2}\right]}\nonumber \\
 & =\frac{\left[4\left(\Delta_{c}-\eta-\delta\right)^{2}+\left(\gamma_{c}\right)^{2}\right]^{2}}{\left[4\left(\Delta_{c}-2\eta-\delta\right)^{2}+\gamma_{c}^{2}\right]\left[4\left(\Delta_{c}-3\eta-\delta\right)^{2}+\left(\gamma_{c}\right)^{2}\right]}.
\end{align}

\end{widetext}

\end{document}